\documentstyle []  {article}

\title {YOUNG STARS AT LARGE DISTANCES FROM THE GALACTIC PLANE: MECHANISMS OF 
 FORMATION}
\author{Christine Allen \footnote{Instituto de Astronomia, UNAM, Apartado, Postal
 70-264 04510, Mexico, D.F. {\it email:} chris@astroscu.unam.mx} \and
T.D.Kinman \footnote{KPNO/NOAO(Operated by AURA, Inc., under contract with the 
NSF), P.O.Box 26732, Tucson, AZ 85726-6732, USA. {\it email:}kinman@noao.edu}}

\date{  }

\begin{document}

\maketitle

\begin{abstract}  
We have collected from the literature data for a list of early-type stars,
situated at large distnces from the galactic plane, for which evidence of youth 
 seems convincing. We discuss two possible formation mechanisms for these stars:
ejection from the plane by dynamical interactions within small clusters, and 
formation away from the plane , via induced shocks created by spiral density
waves. We identify the stars that could be explained by each mechanism.
We conclude that the ejection mechanism could account for about two thirds of
the stars, while a combination of star formation at z = 500-800 pc from the
plane and ejection, can account for 90 percent of the stars. Neither 
mechanism, nor both together, can explain the most extreme examples.
\end{abstract}

\section{Introduction}
There exists an anomalous group of early-type -hence mostly young- stars located 
far 
from the galactic plane, at z-distances ranging from 1 to more than 10 kpc. 
First studied were the A-type stars (Rodgers et al. 1981; Lance 1988). Such 
intermediate-age stars have most recently been studied by Preston and Sneden
(2000) and are likely to be blue stragglers.  
More recently, a number of OB stars have been found, also situated at vertical 
distances of up to several kpc from the plane  (Conlon et al. 1989; Conlon et 
al. 1990; 
Conlon 1992;  Schmidt et al. 1997;  Ringwald et al. 1998).  The shorter 
lifetimes 
of these stars exacerbate the problems of their interpretation in terms of the 
standard picture of star formation and galactic evolution.  The most extreme 
examples,
 if they originated in the plane, must have been ejected with velocities 
surpassing 1000 km/s, which clearly are unrealistically high values.

 Different mechanisms have been put forward to explain the existence of these 
stars, 
either within the conventional view, or postulating star formation in the 
galactic halo
 itself  (see for instance Lance 1988; or Tobin 1991, for extensive reviews).  
These mechanisms range from arguing that  they are misidentified evolved or 
pathological stars, to postulating powerful
ejection mechanisms for young, thin disk stars, or to proposing that they 
were formed in situ, in the galactic halo, from a mixture of gas acquired while 
the Galaxy captured a small satellite galaxy, or from collisions between 
cloudlets, 
or other possibilities.

It has become clear that the anomalous stars are quite a mixed bag themselves, 
and  
that, as a group, they undoubtedly contain some misidentified evolved stars, or 
some Population II stars posing as young B stars.  But the youth of quite a 
number of them seems well established, as shown by accurate determinations of 
surface gravities and colors, high resolution spectral studies, detailed 
abundance determinations,  rotational velocities,  etc.   Among the genuinely 
young OB stars far from the plane, it is also clear that no single mechanism is 
capable of explaining all cases.  Extreme examples, like PG 1002+506, a Be star 
with z $>$ 10 kpc (Ringwald et al. 1998), PG 009+036, a rapidly rotating normal 
B star at z $>$ 5 kpc (Schmidt et al. 1996),  and others, are likely to remain 
puzzling  for the foreseeable future.   Nonetheless, Tobin (1991) concludes that 
dynamical ejection from small clusters in the plane, as envisaged for runaway 
stars by Poveda et al. (1967) and further studied by Gies end Bolton (1986), 
Leonard and Duncan (1988,1990), and others, remains the most likely explanation 
for many of the young stars at large distances from the plane. Clearly, it is 
then of importance to determine for which stars this mechanism is the likely 
explanation

\section{A group of young stars far from the galactic plane}

Although there are many more examples of presumably young stars far from the 
galactic plane scattered in the literature, for the purposes of this work we 
will focus our attention on the relatively homogeneous group of 32 stars studied 
by Conlon et al. (1990).  The evidence for the youth of these stars comes from 
detailed, high resolution abundance studies not only of elements of the CNO 
group, but also of heavier elements.  These determinations allow the authors to 
conclude that their stars are bona fide, normal young B stars, and not evolved, 
intermediate composition stars, or Population II stars mimicking the 
spectroscopic characteristics of early type stars.  

By means of a rudimentary estimation of the times of flight of these stars, 
assuming  they were dynamically ejected from the plane, Conlon et al. conclude 
that ejection is indeed the most likely mechanism to explain the majority of 
them. 

\section{Orbital analysis}

We have obtained improved estimates of the times of flight by numerically 
integrating the 
galactic orbits of these stars.   We have updated the proper motions of the 
Conlon et al. stars using Hipparcos data.  Radial velocities were taken, when 
available, from the Hipparcos Input Catalogue; otherwise, they were calculated 
from the data given by Conlon et al. We used their values for the distances.  We 
then proceeded  to numerically integrate the galactic orbit of each star. The 
galactic potential model of Allen and Santillan (1993)  was used for the 
integration of the orbits.

For each of the stars, times of flight since they left the galactic plane were 
obtained from the orbit computations.  These times of flight were then compared 
with the nuclear lifetimes of disk-composition stars using the models of  
Schaller et al. (1992).  The stellar masses determined by Conlon et al. were 
used.  The results of this analysis are summarized in Table 1.  Successive 
columns contain the Hipparcos number of the star, the time elapsed since it left 
the plane, the mass, the main sequence lifetime, the velocity of ejection from 
the plane, the estimated error in the computed times of flight, the times of 
flight if the stars formed at z = 700 pc, and finally, a code tagging the stars 
that can be explained by the ejection mechanism, or some variants thereof.

Table 1 shows that for 24 out of the 32 stars the times of flight are smaller 
than the nuclear lifetimes.  It is clear, then, that these stars could indeed 
have been ejected from the plane.  They are marked by a 'y' (for yes) in the 
last column of Table 1.  The ejection velocities, also obtained from the orbital 
analysis, are shown in Column 5.  Their values are quite reasonable, and 
compatible with the ejection model. The errors in the times of flight, shown in 
Column 6,  were estimated by computing two additional orbits for each star, with 
the initial conditions modified by the observational errors in distances, proper 
motions and radial velocities. The uncertainties in the times of flight are 
fairly small, and are largely due to the estimated errors in the distances.
 
\section{ Discussion and conclusions }

We have shown that the dynamical ejection mechanism is a plausible alternative 
to explain the majority of the stars in Table 1.  However, for 8 stars the times 
of flight are larger than the nuclear lifetimes, and these stars do not have 
time to reach the z-distances at which they are observed. We could pose the 
question as to whether there are ways to prolong the nuclear lifetimes of 
massive stars.  One obvious possibility is rapid rotation, which will induce 
mixing.   However, models calculated with rotation increase the nuclear 
lifetimes by at most 20 percent (Meynet and Maeder  2000).  So, rotation would 
solve the problem only for three additional stars.  They are marked by an 'r' in 
the last column of Table 1.

Another possibility we can explore is star formation not on the galactic plane, 
but a few hundreds of parsecs above or below it.  Such a mechanism was proposed 
by Martos et al. (2000), and is a result of the passage through the disk of a 
spiral density wave, which can eject sheets of gas to distances of up to 800 pc 
from the plane.  After the spiral density wave passes, the gas will fall back 
onto the plane;  however, in certain cases, Martos et al. showed that conditions 
are favorable for star formation while the ejected gas is still far from the 
plane.  We can envisage that, as occurs in the plane, star formation will result 
not in single stars being born, but rather multiples or small clusters, within 
which the dynamical ejection mechanism could take place.  We would then have 
runaway stars being produced not at z = 0 but at z = 500-800 pc.

Returning to the orbital analysis, we can determine the times of flight not 
since the star left the plane, but since the star left a region situated 700 pc 
above or below the plane, where it could have formed according to the Martos et 
al. scheme.  Such times of flight are, of course, shorter than the times of 
flight from the plane, and could be shorter than the nuclear lifetimes of the 
problem stars.  The stars for which this is the case are marked by an 'f' (for 
birth far from the plane).   There is a total of five stars for which formation 
away from the plane, as in the Martos et al. scheme, would make the ejection 
mechanism a plausible alternative.  

To sum up our results, the last column of Table 1 shows that taking into account 
both the increase in the nuclear lifetimes than can result from stellar 
rotation, and star formation away from the plane, 29 out of the 32 stars can be 
explained by the dynamical ejection mechanism.
This leaves, however, three stars for which another explanation is necessary.

\begin{table}

{\it Table 1}~~~~ RESULTS.

\begin{tabular}{clcllccc}
\hline
{\it Star}&{\it t (flight)} &{\it mass} &{\it m-s life} &{\it v-ej} &{\it $\sigma$(t)} &{\it t 
(700 pc)} &{\it code} \cr
  & 10$^{6}$ y & $M_{\odot}$ & 10$^{6}$ y & km/s & 10$^{6}$ y & 10$^{6}$ y  &  \cr 
\hline

HI001904&51.090&6&\phantom{3}63.1&\phantom{3}93.3&5&&y\\
HI002702&55.530&5&\phantom{3}94.5&\phantom{3}73.7&7&&y\\
HI003812&49.464&9&\phantom{3}26.4&133.8&5&43&no\\
HI006419&26.584&4&164.7&\phantom{3}64.8&5&&y\\
HI011809&12.160&3&352.5&134.4&3&&y\\
HI012320&37.413&3&352.5&\phantom{3}68.8&14&&y\\
HI013800&61.676&6&\phantom{3}63.1&117.1&8&&y\\
HI016130&40.480&4&164.7&\phantom{3}50.7&5&&y\\
HI016466&20.418&5&\phantom{3}94.5&\phantom{3}40.8&4&&y\\
HI016758&21.069&15&\phantom{3}11.6&147.0&2&12&f\\
HI51624AB&12.139&21&\phantom{33}8.0&\phantom{3}51.1&2&0&r,f\\
HI052906&54.068&8&\phantom{3}31.6&171.0&7&48&no\\
HI055051&\phantom{3}5.612&11&\phantom{3}17.6&159.0&1&&y\\
HI055461&39.448&5&\phantom{3}94.5&123.2&3&&y\\
HI058046&13.266&3&352.5&\phantom{3}88.1&5&&y\\
HI059067&48.547&5&\phantom{3}94.5&288.8&2&&y\\
HI059160&14.710&5&\phantom{3}94.5&227.0&1&&y\\
HI059955&21.482&4&164.7&\phantom{3}61.3&3&&y\\
HI060615&37.639&8&\phantom{3}31.6&119.5&4&30&r,f\\
HI070275&13.820&10&\phantom{3}22.4&281.0&1&&y\\
HI079649&24.550&9&\phantom{3}26.4&\phantom{3}84.4&3&&y\\
HI105912&16.585&10&\phantom{3}22.4&140.0&2&&y\\
HI107027&39.930&14&\phantom{3}12.6&212.9&7&33&no\\
HI108215&36.132&5&\phantom{3}94.5&101.0&3&&y\\ 
HI111396&40.060&4&164.7&\phantom{3}64.1&5&&y\\
HI111563&19.605&15&\phantom{3}11.6&228.0&2&14&f\\
HI112790&60.498&4&164.7&\phantom{3}88.1&8&&y\\ 
HI113735&13.903&9&\phantom{3}26.4&161.0&1&&y\\ 
HI114569&\phantom{3}8.316&6&\phantom{3}63.1&350.7&1&&y\\
HI114690&12.015&19&\phantom{33}8.6&157.4&1&8&r,f\\ 
HI115347&21.675&9&\phantom{3}26.4&\phantom{3}58.1&4&&y\\
HI115729&24.607&8&\phantom{3}31.6&\phantom{3}76.7&3&&y\\ 

\hline
\end{tabular}

\end{table}

\end{document}